\documentclass[]{spie}  

\usepackage{amsmath,amsfonts,amssymb}
\usepackage{graphicx}
\usepackage[colorlinks=true, allcolors=blue]{hyperref}
\usepackage{siunitx}
\usepackage{xcolor}

\newcommand\lcdm{$\Lambda$CDM}

\DeclareSIUnit{\beli}{Bi}
\DeclareSIUnit{\dBi}{\deci\beli}

\title{Modelling ground pickup for microwave telescopes}

\author[1,*]{Alexandre E. Adler}
\author[2]{Adriaan J. Duivenvoorden}
\author[1]{Jon E. Gudmundsson}

\affil[1]{{The Oskar Klein Centre, Department of Physics, Stockholm University, SE-106 91 Stockholm, Sweden}}
\affil[2]{Joseph Henry Laboratories of Physics, Jadwin Hall, Princeton University, Princeton, NJ, USA 08544}
\authorinfo{*Send correspondence to Alexandre E. Adler, E-mail: alexandre.adler@fysik.su.se}

\begin{document} 
\maketitle

\begin{abstract}
    Microwave telescopes require an ever-increasing control of experimental systematics in their quest to measure the Cosmic Microwave Background (CMB) to exquisite levels of precision. 
	One important systematic for ground and balloon-borne experiments is ground pickup, where beam sidelobes detect the thermal emission of the much warmer ground while the main beam is scanning the sky.
	This generates scan-synchronous noise in experiment timestreams, which is difficult to filter out without also deleting some of the signal from the sky.
	Therefore, efficient modelling of pickup can help guide the design of experiments and of analysis pipelines.
	In this work, we present an extension to the {\sc beamconv} algorithm that enables us to generate time-ordered data (TOD) from beam-convolved sky and ground maps simultaneously. 
	We simulate ground pickup for both a ground-based experiment and a telescope attached to a stratospheric balloon. 
	Ground templates for the balloon experiment are obtained by re-projecting satellite maps of the Earth's microwave emission.\end{abstract}


\section{INTRODUCTION}
\label{sec:intro}  
Studies of the microwave sky are crucial for cosmology. 
The Cosmic Microwave Background (CMB) is the oldest light in the Universe: it has been travelling towards Earth since roughly 380,000 years after the Big Bang.
We observe the CMB as a nearly isotropic field of blackbody radiation of $\sim\SI{2.7}{\kelvin}$\cite{Fixsen2009}. 
Small spatial fluctuations (anisotropies) in the blackbody temperature, of the order of a few parts per $10^4$ have been measured by several ground\cite{Kovac1998, ACTPol2016, POLARBEAR2010, BICEP2}, balloon-borne\cite{Crill2003, Hanany2000, Nagy2017b} and space-based experiments\cite{Smoot92, WMAP2013, Planck2018_overview}. 
These, in turn, lead to stringent constraints on the history and evolution of the Universe, both before and after the CMB's emission.
In fact, the CMB underpins the consensus cosmological model, \lcdm. 
Beyond the temperature anisotropies, the CMB is also partly polarised \cite{Kovac2002}. 
This polarisation is weaker than the temperature anisotropies, thus detecting it has proved more challenging. 
The polarisation state of electromagnetic radiation can be expressed using the Stokes parameters: total intensity $I$, the two linear polarisation directions $Q$ and $U$, and circular polarisation $V$. 
Together, they form the Stokes vector $(I,Q,U,V)$. 
So far no circular polarisation of cosmological origin has been observed\cite{Nagy2017, Padilla2020}.
The $Q$ and $U$ polarisation can be combined into two rotationally invariant (scalar) quantities: the parity-even $E$-mode field and the parity-odd $B$-mode field.

$E$ and $B$-modes are produced by a variety of processes in the CMB. 
For the sake of brevity and clarity, we refer the reader to \citenum{Komatsu2022}, a recent review of the topic. 
We will limit ourselves to stating that the detection of primordial CMB $B$-mode polarisation would be strong evidence in favor of inflation\cite{Crittenden93}, a proposed period of extremely rapid expansion in the first instants of the Universe.
Finding solid evidence that inflation happened would have enormous consequences for cosmology and high-energy physics.
It is therefore no surprise that many experiments are aiming to detect this polarisation\cite{CMBS4_2017_technology_book, LiteBIRD2018}. 

Advances in detector technology, cryogenics, and optical design have made microwave detectors more compact and less noisy. 
The extremely challenging nature of the measurement means these experiments need an exquisite control over possible sources of systematic error. 
The optical design, the scanning strategy, the electronics and readout chain, are all chosen with the need to minimize systematics in mind.
For that reason, simulations of possible systematics are a key aspect of the design process. 

A persistent systematic for ground-based and balloon-borne experiments is scan-synchronous noise (SSN). 
Each detector in a CMB experiment generates a timestream as an output. 
CMB telescopes usually scan back and forth in azimuth at a fixed or near-fixed elevation.
A contaminant fixed in azimuth will thus produce spurious correlations in the noise over long periods of time as the telescope comes back to the same azimuth repeatedly.

The leading source of SSN for ground and balloon experiments is ground pickup. 
It results from the interaction between the telescope's beam $B(\theta, \phi)$ and the ground surrounding the telescope. 
The total ground pickup is the convolution between the beam and the ground.
The beam for a telescope is usually strongly directional. 
It is decomposed into the main beam, which concentrates the quasi-totality of the power, and sidelobes, which are much weaker but extend to large angles from the beam's maximum. 
While the sidelobes are typically very weak, the ground is extremely warm compared to the CMB anisotropies. 
As a consequence, the amplitude of the coupling between telescope sidelobes and the ground can generate noise comparable in intensity to the sky signal. 
As the ground is fixed in the telescope's frame of reference, that noise falls under the umbrella of SSN. 

Current ground based experiments typically suppress the amount of SSN that is propagated to the science products (sky maps, angular power spectra, cosmological parameter estimations) by effectively high-pass filtering the timestreams\cite{Poletti2017}. 
If the filter frequency is higher than the frequency at which the telescope comes back at the same (az, el) position, then the correlations in the timestream associated with telescope motion will be suppressed. 
However, SSN is not the only cause for long-period correlations in timestreams. 
Indeed, large scale CMB anisotropies will lead to low-period features in experimental timestreams. 
Then the high-pass filter will also reduce what the experiment can learn about the large angular scales. 
The only observational window on $B$-mode polarisation from primordial gravitational waves is expected to be at multipoles of $\ell\sim 100$ (roughly \SI{1.8}{\degree}) or smaller. 
It is clear that a CMB experiment interested in probing these large angular scales will want to minimize ground pickup so it can be less aggressive with its timestream filtering.

As mentioned earlier, experiments rely on simulations to guide their design choices. 
In this proceeding, we present an effort to simulate timestreams for a sky signal contaminated by ground pickup. 
We present what is, to our knowledge, the first open source library to generate timestreams for the beam-convolved ground and sky simultaneously.
In particular, we generate templates of ground emission for two types of telescopes. 
The first is a telescope situated in the heights of the Atacama Desert, near where the Atacama Cosmology Telescope and the future Simons Observatory are located.
The second is a balloon-borne telescope whose position and altitude can change from day to day. 
As the balloon finds itself flying over a wide variety of terrains, we need to simulate ground templates for many different areas.
To that end, we re-project satellite maps of the Earth's brightness temperature. 
The templates are described in Section~\ref{sec:ground}.
In both cases, we model the instrument as a two-lens refractor. 
Beams are simulated for 50 detectors. 
This constitutes Section~\ref{sec:telescope}.
Beams and ground template are then convolved together using the fast convolution algorithm {\sc beamconv}\cite{Duivenvoorden2018} to generate timestreams following a set scan strategy. 
This constitutes Section~\ref{sec:beamconv}
We present the results in Section~\ref{sec:results}, and discuss the next steps in Section~\ref{sec:conclusion}.
\section{GROUND MODELS}
\label{sec:ground}
\subsection{Modelling of the Atacama Desert}\label{ssec:gata}
\begin{figure}
    \centering
    \includegraphics[width=\textwidth]{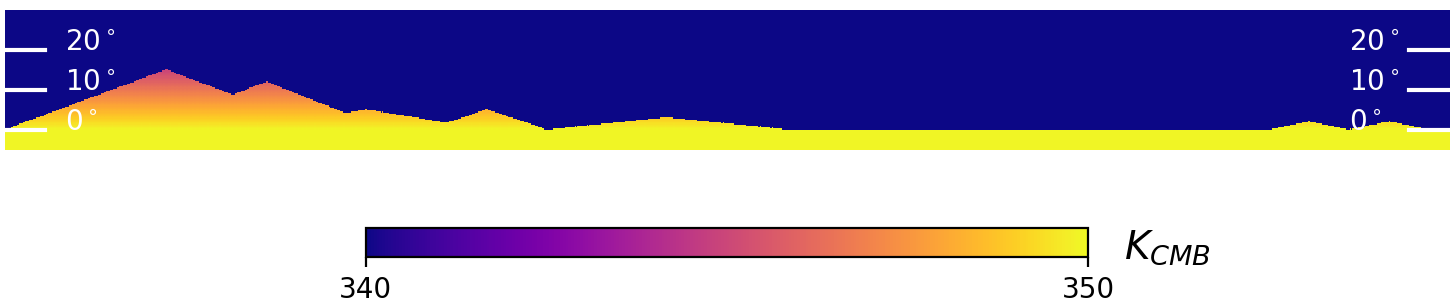}
    \caption{Ground template created from site photographs and local temperature estimates, converted to CMB brightness temperature units at \SI{95}{\giga\hertz}. 
    The temperature of the sky is set at \SI{0}{\kelvin}.
    The highest peak, Cerro Toco, is \SI{400}{\metre} higher than the site and a few kilometers away.}
    \label{fig:terrainphoto}
\end{figure}

We place our telescope on the slopes of Cerro Toco, at the location of the current Atacama Cosmology Telescope\cite{ACTPol2016}. 
A panoramic view of the surrounding terrain can be seen in Figure \ref{fig:terrainphoto}. 
In that picture, a number of mountains of various heights rise over the horizon. 
We simplify that landscape down to a number of triangles. 
The ground around the experiment up to the horizon, is assumed to be a blackbody with a temperature of \SI{300}{\kelvin}, which translates to a brightness temperature of $\SI{3.5e8}{\micro\kelvin}_{\textrm{CMB}}$.
The emission temperature of the mountains decreases with altitude at a rate of \SI{1}{\celsius} per \SI{100}{\metre}.
\subsection{Modelling the ground for a balloon-borne experiment}
Temperature does not convert to brightness temperature uniformly for different types of terrain. 
Since balloons will drift over enormous distances and can take very different paths, we cannot reasonably expect to model the temperature and the terrain analytically like in Section~\ref{ssec:gata}.
Fortunately, there exists actual data of the Earth's brightness temperature in the microwave band. 
Several of the US Defense Meteorological Satellite Program (DMSP) satellites carry an instrument called SSMIS\cite{Bommarito1993, Kunkee2008}, which stands for Special Sensor Microwave Imager/Sounder. 
SSMIS measures the brightness temperature of the ground in several bands\cite{Brodzik2016}.
We extract maps of emission in the \SI{91}{\giga\hertz} band.
By combining the different scan patterns and regions to generate near-complete daily maps of the Earth's microwave temperature. 
The maps are projected following the {\sc healpix}\cite{Healpix2005} pixellization convention, while assuming that the Earth is spherical.

We then need to transform these Earth maps into ({\textit Az-el}) maps from the telescope's point of view. 
An observer at an altitude $z$ above the North Pole of a sphere will see the surface up to a co-latitude of $\alpha_{fov}$:
\begin{equation}
    \alpha_{fov} = \cos^{-1}\left({\frac{R}{R+z}}\right) \: ,
\end{equation}
where $R$ is the sphere's radius.
A point on the sphere with latitude $\lambda$ and longitude $l$, within the spherical cap of angular radius $\alpha_{fov}$, will get projected in a spherical coordinate system centered on the observer:
\begin{align}
        \theta =& \pi - \tan^{-1}\left(\frac{\sin(\pi/2 - \lambda)}{(1-\cos(\pi/2 - \lambda)) + \frac{z}{R}}\right) \: ,\label{eq:theta} \\
         \phi = &\pi - l \: ,\label{eq:phi}
\end{align}
with $(\theta, \phi)$ the co-latitude and longitude in that new coordinate system.
The choice of zero longitude is arbitrary, we impose that choice such that if our observer was instead at a latitude $\lambda = \pi/2- \epsilon$, $\epsilon \ll 1$ and longitude \SI{0}{\degree}, $\phi=0$ would point towards the North Pole of the sphere. 

\begin{figure}
    \centering
    \includegraphics[width=.85\textwidth]{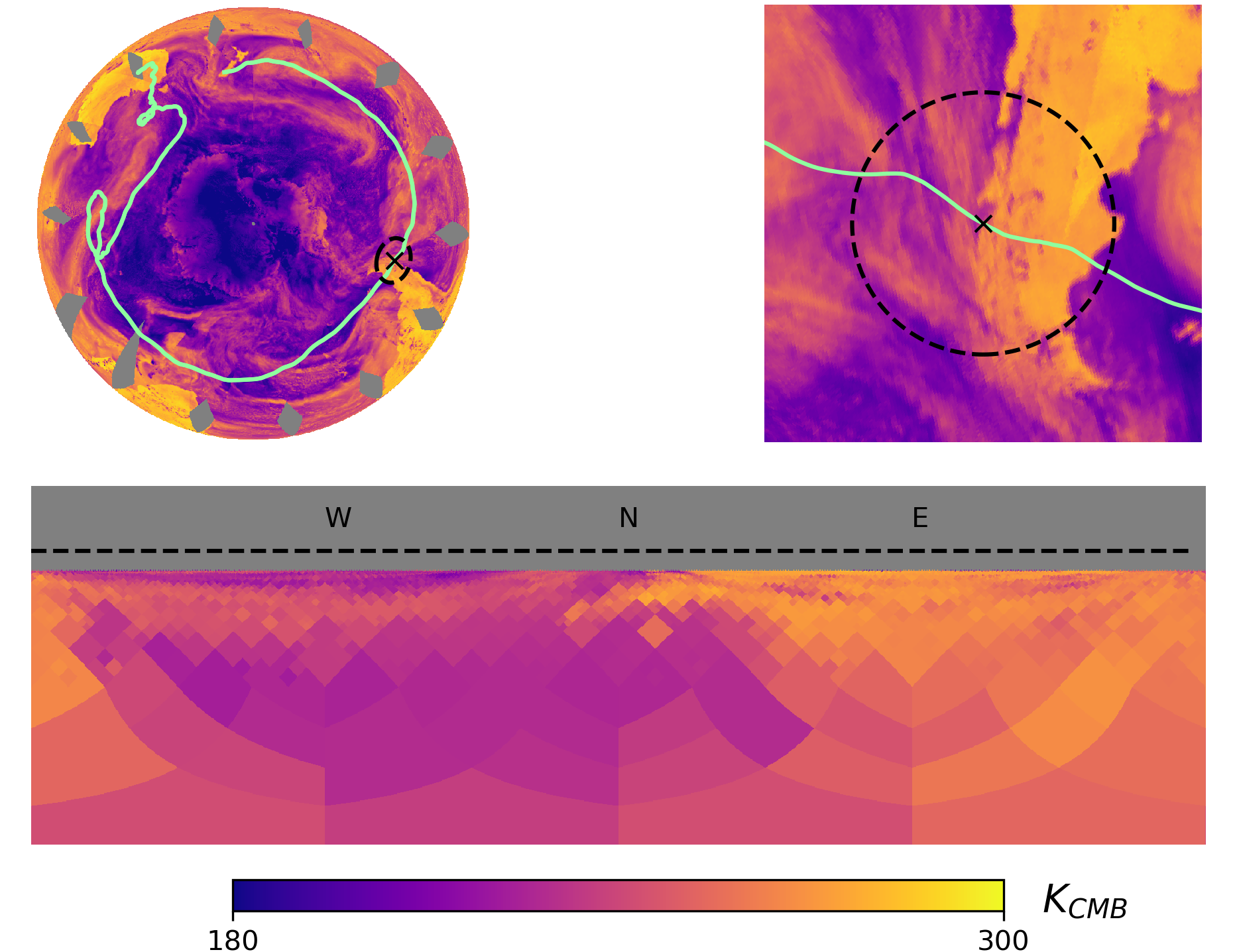}
    \caption{Top left: a map of microwave emission at \SI{91}{\giga\hertz} from the Southern Hemisphere for a given day, obtained from SSMIS data. 
    The path of the balloon is overlaid in green. Top right, example of a balloon's field of view (black circle) as it passes over the coast of Chile at an altitude of \SI{35}{\kilo\metre}.
    Bottom panel: projection of the ground in the telescope's frame of reference. 
    The dashed line indicates zero elevation, the horizon is \SI{6}{\degree} below that. Letters denote cardinal directions.}
    \label{fig:balloontemplate}
\end{figure}

Given a position $(\lambda, l)$ and an altitude $z$ for our telescope, we rotate the Earth template such that the $(\lambda, l)$ corresponds to the pole, and then project all pixels within $\alpha_{fov}$ using Equations~\ref{eq:theta} and \ref{eq:phi}. 
The resulting ground template for the telescope is then infilled up to the horizon line. 
A depiction of the process can be seen in Figure~\ref{fig:balloontemplate}.
We make the code used to generate the projection available in a public Github repository. \footnote{\href{https://github.com/AERAdler/thefloorisdata/}{https://github.com/AERAdler/thefloorisdata}}
\section{THE TELESCOPE}
\label{sec:telescope}
Our telescope is based on the two-lens silicon refractor design presented in \citenum{Gudmundsson2020}, which we already worked with in a previous proceeding\cite{Adler2020}. 
As we are focused on the large angular scales, the design is optimized to have a wide diffraction-limited field of view (DLFOV). 
The telescope should have consistent performance from 30 to \SI{350}{\giga\hertz}, since multi-frequency observations are needed to disentangle foregrounds. 
The two lenses are roughly \SI{310}{\milli\metre} in diameter, with a sag of \SI{16}{\milli\metre} on the primary. 
The axial distance from primary to secondary is \SI{450}{\milli\metre}, and from secondary to the focal plane \SI{137}{\milli\metre}. 
Accounting for lens thickness, this results in a compact telescope roughly \SI{625}{\milli\metre} in length. 
The DLFOV is \SI{30}{\degree} up to \SI{350}{\giga\hertz}.
The focal plane has a diameter of \SI{250}{\milli\metre}, therefore the plate scale is $\sim\SI{0.12}{\degree\per\milli\metre}$.
With a pitch size of \SI{6}{\milli\metre}, this means around $\sim 1300$ pixels could be fit on that focal plane. 

The design is simulated with the optical modelling software TICRA-Tools\cite{TicraTools}, formerly GRASP. 
Using physical optics (PO) and the Physical Theory of Diffraction (PTD), we can simulate the propagation of radiation in the time-reversed sense, that is, from the focal plane to the far-field. 
We generate far-field beams for a given horn on the focal plane, modelling the radiation propagating from the horn through the secondary and the primary lenses. 
Scattering by a vacuum window in front of the primary lens and internal reflection in the lenses are approximated as a Lambertian source placed just in front of the primary. 
The total power of the Lambertian source is set to be \SI{1}{\percent} of the total beam power, in line with theory\cite{Ruze66}. 
The effect of an absorbing forebaffle outside of the optics is approximated by an aperture in an infinite screen situated at the forebaffle's end. 
The Lambertian source and the beam exiting the primary lens are propagated through the aperture and to the far-field.
We do not attempt to simulate the cryostat that encloses the lenses and the focal plane. 

\begin{figure}\label{fig:beam}
    \centering
    \includegraphics[width=\textwidth]{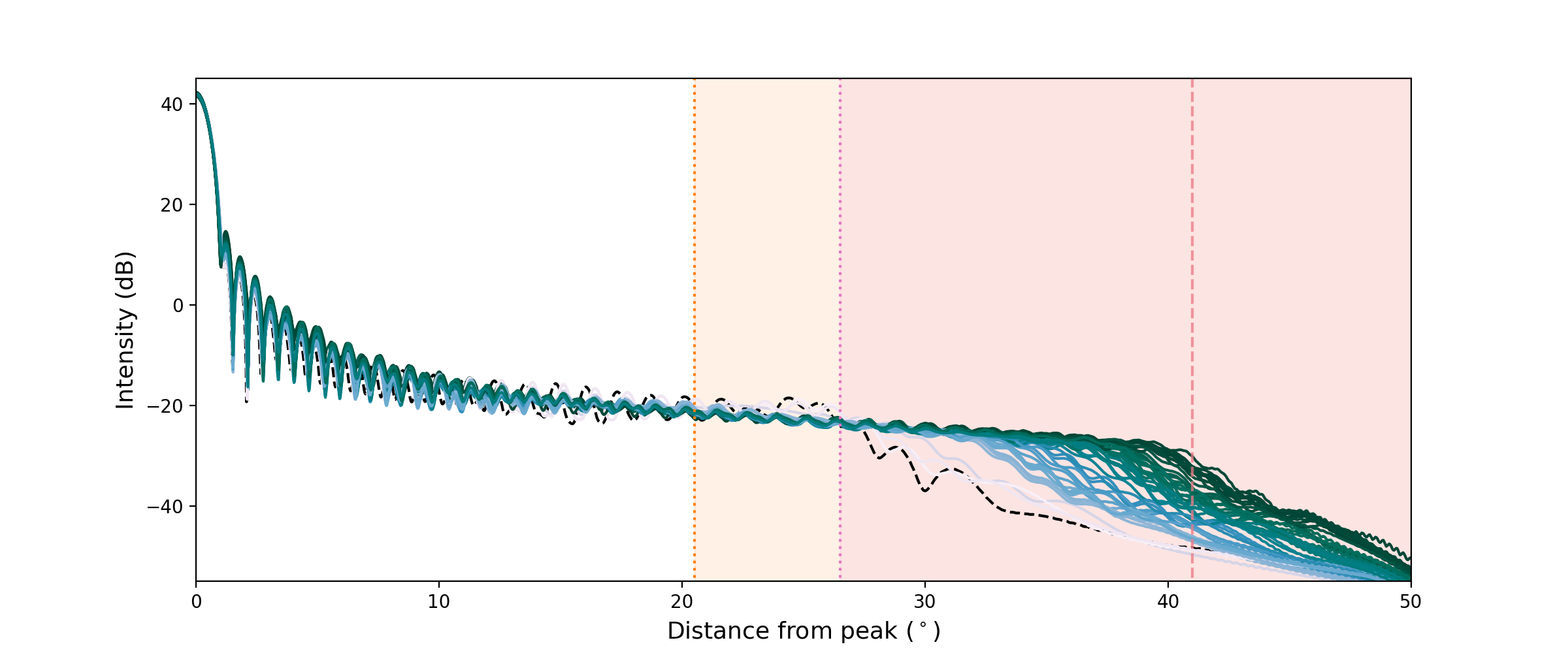}
    \caption{Azimuthally-averaged Stokes I far-field beam profile for all 50 detectors at \SI{95}{\giga\hertz}. 
    A darker color indicates greater distance from the center of the focal plane to the detector, up to \SI{14.5}{\degree} off-axis.
    The dashed black line represents the far-field beam profile for an on-axis detector.
    We distinguish a 46.5' full-width half maximum main lobe and a much weaker sidelobe.
    The orange outline shows parts of a beam that can intersect the ground for the ground-based experiment, with the vertical dotted line marking when a beam on the edge of the focal plane will start to see Cerro Toco.
    The pink outline represents the same situation for the balloon-borne telescope.
    The dashed pink line shows when a beam at the  center of the focal plane will start to detect the ground. 
    By coincidence, it is the same value for both telescopes.}
\end{figure}

We select 50 pairs of (x,y) positions at random on the focal plane and conduct PO simulations for each of them at \SI{95}{\giga\hertz}. 
The far-field beams are sampled on two grids: a $\SI{3}{\degree}\times\SI{3}{\degree}$ grid samples the main beam finely, while a coarser $\SI{60}{\degree}\times\SI{60}{\degree}$ grid enables us to probe the sidelobes. 
The co- and cross-polar beams in the TICRA-Tools output are converted to {\sc healpix} maps of the Stokes parameters $(I, Q, U)$. 
Figure~\ref{fig:beam} shows the azimuthally-averaged Stokes \textit{I} beam profiles at \SI{95}{\giga\hertz}.
Beams that are further off-axis will have comparatively more extended sidelobes as the main beam's centre moves relative to the centre of the scattering.  

\section{CONVOLUTION, TIMESTREAMS, MAPS}
\label{sec:beamconv}
The signal measured by a detector pointing at celestial coordinates $(ra, dec)$ is the convolution of the sky signal and the beam over the full $ \SI{4\pi}{\steradian}$ of the sky. 
CMB experiments typically sample their signal several times per second, and can observe for several weeks or month. 
Therefore, convolution in real space for each of these measurements is computationally very costly.
However, convolution in real space is equivalent to multiplication in harmonic space.
The beam and sky maps can be decomposed into the spherical harmonic base, giving us two set of coefficients ($b_{\ell m}$ and $a_{\ell m}$ respectively).
Multiplying these together is equivalent to convolving the sky and the beam in real space\cite{wandelt_2001, challinor_2000, Prezeau2010}. 

This is the motivation behind {\sc beamconv}, a python package that can generate streams of time-ordered data (TOD) when given the spherical harmonic coefficients and an observation schedule. 
Originally, {\sc beamconv} was developed to probe the effect of various beam systematics on our ability to reconstruct accurate sky maps and CMB power spectra.
{\sc beamconv} uses spin-weighted spherical harmonics (SWSH) to accurately capture the polarised nature of the beam and the sky.
At a given $(ra, dec)$, the data model simulated by {\sc beamconv} is\cite{Duivenvoorden2021}:

\begin{equation}
d =  \!\sum_{\ell, m, s} \Big\{ b^{\widetilde{I}}_{\ell s}\! a^{I}_{\ell m} \! + b^{\widetilde{V}}_{\ell s}\! a^{V}_{\ell m} 
\! + \frac{1}{2} \Big[ {}_{-2}b^{\widetilde{P}}_{\ell s}\! {}_{2}a^{P}_{\ell m}  + {}_{2}b^{\widetilde{P}}_{\ell s}\! {}_{-2}a^{P}_{\ell m} \Big] \Big\} 
 \times  \sqrt{\frac{4 \pi}{2 \ell + 1}} \mathrm{e}^{- \mathrm{i} s \psi} {}_s Y_{\ell m} (ra, dec) ,
\end{equation}
with $d$ the measured time-ordered signal, $P=Q+iU$ the complex form of the polarisation vector, ${}_s a^{X}_{\ell m}$ the SWSH coefficient for sky component $X$ and $\psi$ the position angle of the detector. 
In this work, we modified the data model, replacing the sky SWSH coefficients with those of the ground and evaluating the SWSH in $(az, el)$ coordinates, thereby generating a ground pickup signal that gets added to the normal {\sc beamconv} timestreams.

\begin{figure}
    \centering
    \includegraphics[width=\textwidth]{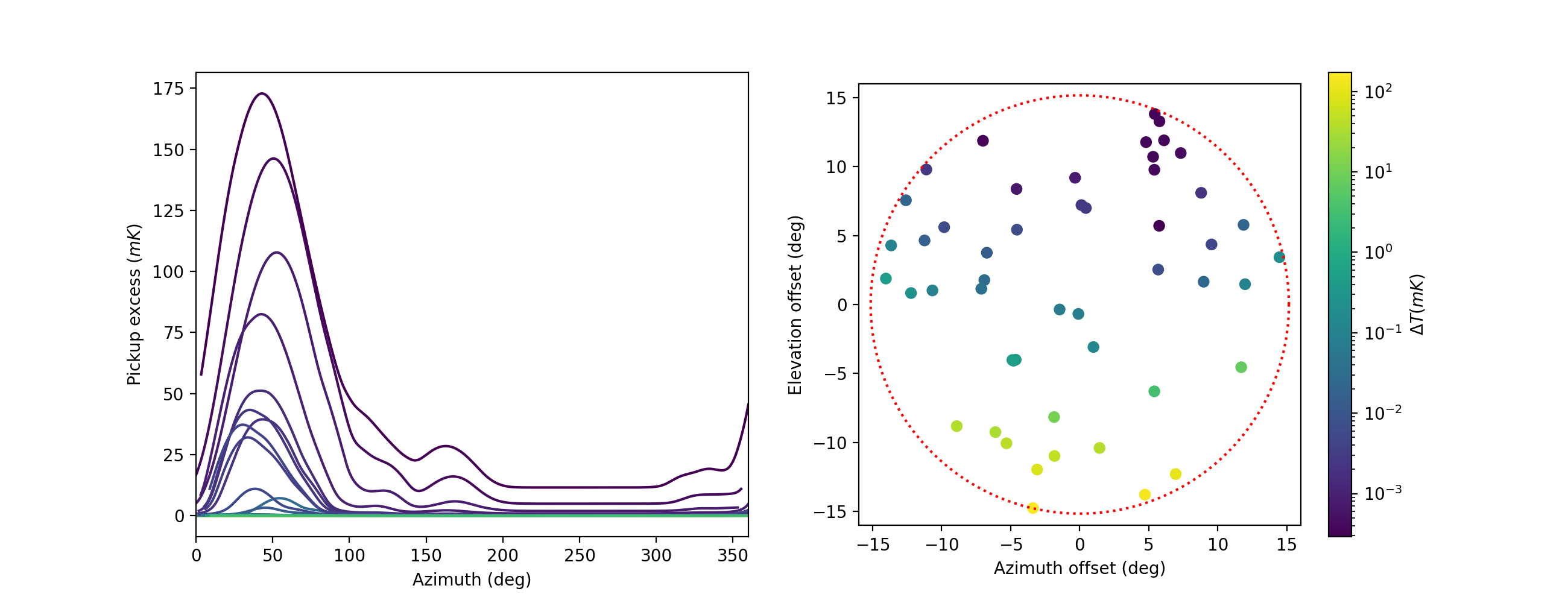}
    \caption{Left: Difference between ground-contaminated and pure sky timestreams for a full rotation in azimuth at a fixed boresight elevation of \SI{55}{\degree}.
    The x-axis gives the azimuth at which a particular detector is pointing.
    Different colors denote elevation offsets in the range [\SI{-14.5}{\degree}, \SI{0}{\degree}], with dark blue indicating the timestream of a detector pointing \SI{14.5}{\degree} lower than the boresight.
    All detectors pointing higher than the boresight have had their timestream colored in yellow.
    Right: Pointing of the 50 detector pairs: their positions have been generated at random. 
    The color of each dot indicates the maximum difference between the two timestreams.}
    \label{fig:timestream}
\end{figure}

To demonstrate the algorithm, we first perform a constant-elevation-scan with our 50 detectors over all \SI{360}{\degree} of azimuth with and without the Atacama-like ground template. 
In Figure~\ref{fig:timestream}, we see the difference between these two timestreams for the 'A' detector of each detector pair. 
The outline of the main elevation features are very visible for the several of the timestreams, with detectors that point lower in elevation picking up a larger ground signal. 
A representation of the focal plane is visible as the right panel in Figure~\ref{fig:timestream}.
It is also interesting to notice that the azimuth of the main feature, which corresponds to pickup of the Cerro Toco peak, varies from timestream to timestream. 
Complexities of the beam model will therefore impact both the amplitude of pickup and its azimuthal dependence. 

All ground-based CMB experiments filter their TOD in some way. 
Depending on the expected noise, the best filtering procedure might differ.
In simulations and previous experiments, high-pass, pass-band\cite{Duivenvoorden2021}, or polynomial\cite{SPIDER2022} filters have been applied to TOD.
In our case, we implement a first-order Butterworth high-pass filter, with the cutoff frequency set to \SI{0.1}{\hertz}.
The TOD is then combined with detector pointing information and binned to create $(I,Q,U)$ sky maps using a naive map-making algorithm. 
By differentiating output maps that follow the same scan strategy, we can gauge the effects of different modelling choices. 
In particular, we can look at the difference maps between scans with and without the presence of a ground template, or how well a particular filter reduces pickup.

\section{RESULTS}\label{sec:results}
Each of our two telescopes has its own scan strategy.
The ground telescope's constant elevation scan probes a smaller area of the sky, but very deep, while the balloon experiment's scan strategy covers more than half of the celestial sphere.

\begin{figure}
    \centering
    \includegraphics[width=\textwidth]{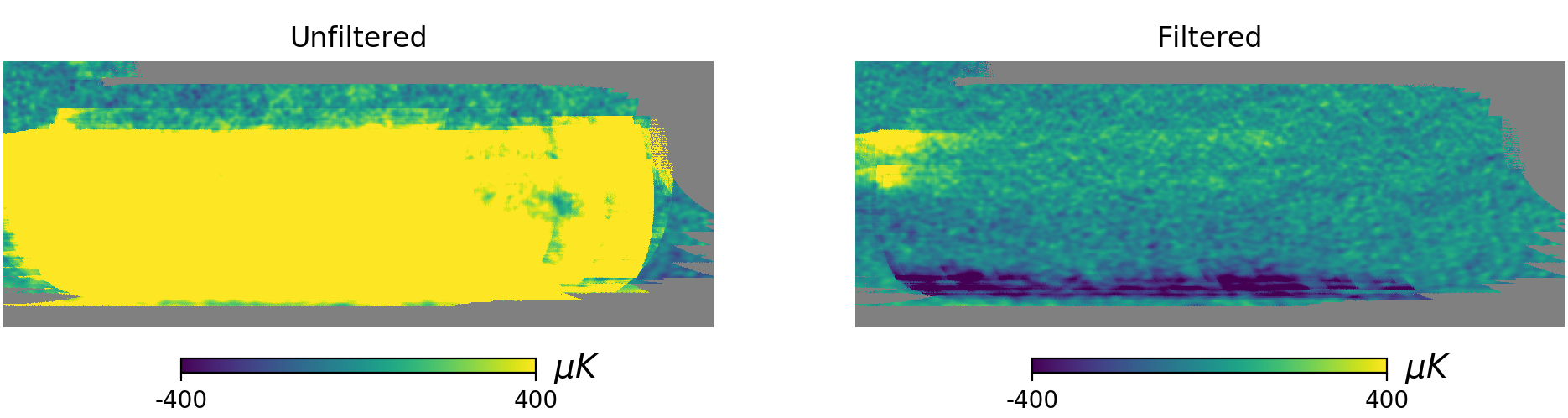}
    \caption{Binned Stokes I maps for the ground-based telescope in the region $-50\leq \textrm{ra}\leq110$, $-70\leq\textrm{dec}\leq-10$.  
            Left panel: un-filtered timestreams get binned into a map dominated by ground pickup.
            Towards the top of the map, a region with less pickup shows the CMB signal.
            Right panel: applying the high-pass filter on the timestreams gets rid of the majority of the pickup, but also suppresses large-scale CMB temperature anisotropies. }
    \label{fig:atacama_map}
\end{figure}

The ground-based telescope performs a constant-elevation scan, alternating between two patches of the sky separated by the Galactic plane which we typically call the "Southern" and "Northern" patches.
The telescope slews in azimuth between two limits at a fixed elevation at a rate of \SI{1}{\degree\per\second}, going back and forth.
Every ten minutes, the telescope moves in elevation to track the motion of the original patch center. 
Boresight elevation is constrained between 50 and \SI{62}{\degree}.
The telescope samples the sky at $\SI{50}{\hertz}$ for 31 days.
We show results for the Southern Patch in Figure~\ref{fig:atacama_map}. 
There is a large amount of pickup due to the ground, which the filter is partly successful at cutting out. 

\begin{figure}
    \centering
    \includegraphics[width=\textwidth]{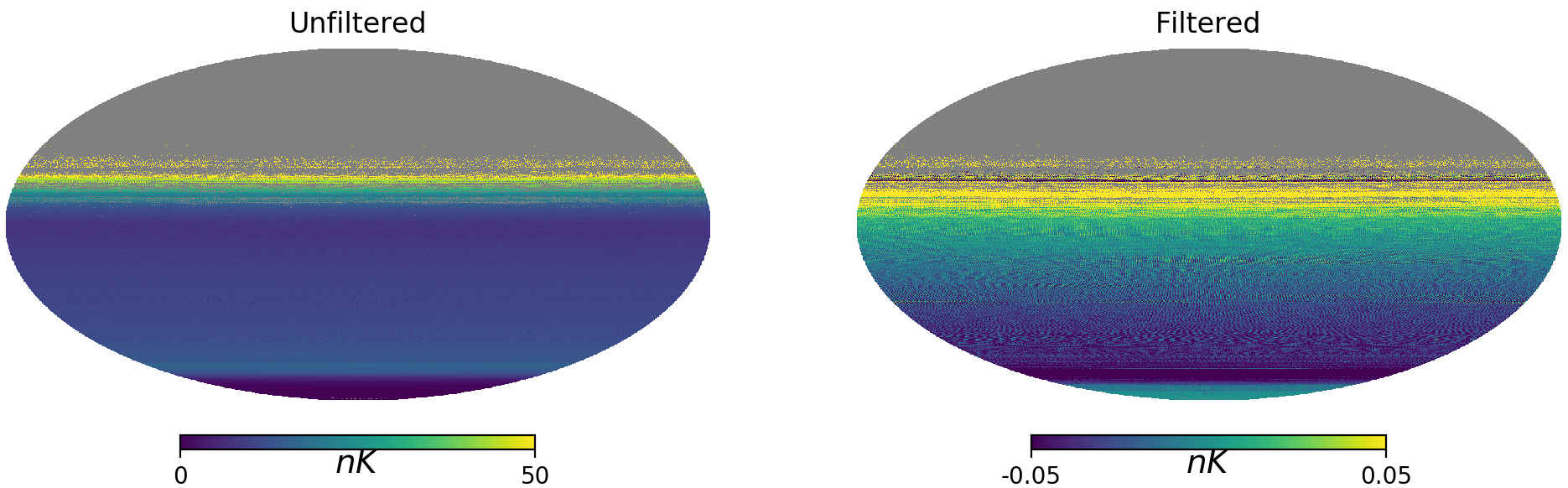}
    \caption{Difference in Stokes I maps between scans with and without ground SSN for the balloon-borne telescope's observations. 
            The pickup is globally quite weak, and concentrated towards the northern-most declinations sampled by the telescope.}
    \label{fig:balloon_map}
\end{figure}

The balloon-borne telescope drifts beneath the balloon at an altitude of \SI{35}{\kilo\metre} above sea level, following the path depicted in the top-left plot of Figure~\ref{fig:balloontemplate}. 
The position is updated daily for a month. 
It rotates continuously in azimuth at a speed of \SI{30}{\degree\per\second} at a fixed elevation of \SI{35}{\degree}. 
Every hour, the direction of rotation is reversed. 
We make daily maps that are then co-added to give the map for the whole simulated mission. 
The results can be seen in Figure~\ref{fig:balloon_map}.
The pickup is concentrated at positive declinations, reflecting that these can only be accessed from positions lower on the horizon in the Southern Hemisphere. 
It is much weaker than in the ground-based simulation, in line with the observation in Figure~\ref{fig:beam} that a smaller fraction of each beam can pickup the ground emission.

\section{CONCLUSION}
\label{sec:conclusion}
We have presented an extension to the {\sc beamconv} code than enables the generation of time-ordered data from beam-convolved ground maps. 
We demonstrated that new extension by simulating two experimental setups, a ground-based telescope picking up thermal emission from nearby terrain and an instrument suspended from a balloon drifting through the stratosphere. 
In order to generate the second setup's ground templates, we also wrote a small separate program to project daily maps of microwave emission from the SSMIS satellite into ground templates in the telescope's frame of reference. 
Our ground templates make a number of simplifying assumptions, like the absence of a ground screen for the ground-based telescope or ignoring solar reflections and the balloon itself for the airborne instrument. 
While our beams were generated from a fictional telescope design, our framework can easily be adapted to simulate scan-synchronous noise for real current and future experiments. 
In particular, we aim to use all of {\sc beamconv}'s current capabilities to study systematics effects for a future Ultra Long Duration Balloon CMB experiment.
On top of pickup effects, we will include half-wave plate and beam non-idealities.
\acknowledgments 
 
The authors would like to thank Steven Benton, Kevin Crowley, Nadia Dachlythra and Reijo Keskitalo for productive discussions. Alexandre Adler's participation to this SPIE conference is funded by the Birger and Gurli Grundstroms research grant fund.

\bibliography{main} 
\bibliographystyle{spiebib} 

\end{document}